\documentstyle[pra,aps,amssymb,amsmath,epsf,multicol,psfrag]{revtex}

\def\unity{\mbox{\small 1} \!\! \mbox{1}}

\begin{document}

\title{Detection devices in entanglement-based optical state preparation}
\author{Pieter Kok and Samuel L.\ Braunstein}
\address{Informatics, University of Wales, Bangor, LL57 1UT, UK}

\maketitle

\begin{abstract}
 We study the use of detection devices in entanglement-based state preparation.
 In particular we consider optical detection devices such as single-photon
 sensitivity detectors, single-photon resolution detectors and detector
 cascades (with an emphasis on the performance of realistic detectors). We 
 develop an extensive theory for the use of these devices. In
 entanglement-based state preparation we perform measurements on subsystems,
 and we therefore need precise bounds on the distinguishability of these
 measurements (this is fundamentally different from, e.g., tomography,
 where an ensemble of identical states is used to determine probability
 distributions, etc.). To this end, we introduce the confidence of preparation,
 which may also be used to quantify the performance of detection devices in 
 entanglement-based preparation. We give a general expression for detector  
 cascades of arbitrary size for the detection up to two photons. We show that,
 contrary to the general belief, cascading does not give a practical advantage 
 over detectors with single-photon resolution in entanglement-based state
 preparation. 
\end{abstract}

PACS number(s): 42.50.Ar

\begin{multicols}{2}
The accurate creation of quantum states is important to many applications in,
for example, quantum computation and information \cite{bouwmeester00,peres95}. 
One method of state preparation is to entangle two systems and subsequently 
perform a so-called {\em conditional} measurement on one subsystem: depending 
on the measurement outcome the undetected subsystem is `prepared' (collapsed) 
into a particular predetermined state (see also Rubin \cite{rubin00}). 
Considerable progress has been made using this method in the creation of {\em 
optical} quantum states \cite{vogel93,harel96,schleich97,dariano00,dakna99} 
and in the creation of three-photon polarisation entanglement 
\cite{zeilinger97}. Optical entanglement sources include, for example, 
cross-Kerr media \cite{dariano00} or the mixing of states at beam-splitters 
\cite{dakna99}. In general, the quality of this entanglement-based state 
preparation strongly depends on the details of the conditional measurement. 

In this paper we study the effect of {\em realistic} (photo-) detectors on the 
state preparation process. To this end we introduce the concept of the {\em 
confidence} of preparation in Sec.\ \ref{confidence}. This measure does not 
only quantify the `quality' of the state preparation process, but it also 
allows us to compare different types of detection devices. In Sec.\ \ref{odd} 
we discuss  the distinction between single-photon sensitivity and single-photon
resolution detectors. The statistics of detector cascading with single-photon
sensitivity detectors is studied in Sec.\ \ref{sec:nports} and Sec.\ 
\ref{compare} makes a numerical comparison between such detector cascades and
single-photon resolution detectors.

Let's consider entanglement-based state preparation \cite{rubin00} (not 
necessarily restricted to quantum optics). We want to prepare a {\em single} 
(pure) state $|\phi\rangle$ by means of some entanglement-based process, and 
we want the resulting state $\rho$ to be as `close' to $|\phi\rangle$ as 
possible. A measure of resemblance between states is given by the fidelity $F$ 
\cite{fuchs96}:
\begin{equation}
 F = {\rm Tr} [\rho |\phi\rangle\langle\phi|]\; .
\end{equation}
The quality of a state preparation process can therefore be measured by the
fidelity. When $F=1$, the process gives exactly $|\phi\rangle$ and when $F=0$,
the prepared state is orthogonal to $|\phi\rangle$. In practice, the fidelity 
will not reach these extreme measures, but will lie between 0 and 1.

For example, if we want to prepare a single-photon state $|1\rangle$ we can 
use the following process: a parametric down-converter creates a state 
$|\psi\rangle_{ab}$ on two spatial modes $a$ and $b$:
\begin{equation}
 |\psi\rangle_{ab} \propto |0\rangle_a |0\rangle_b +\xi|1\rangle_a |1\rangle_b 
 + O(\xi^2)\; ,
\end{equation}
where $|0\rangle$ denotes the vacuum state and we assume $\xi\ll 1$. The 
higher order terms (included in $O(\xi^2)$) consist of states with more than 
one photon. We now place a photo-detector in mode $a$, which `clicks' when it 
sees one or more photons (typically, standard detectors can see single photons,
but fail to distinguish between one and two photons). Conditioned on such a 
click, mode $b$ will be in a state
\begin{equation}
 \rho \propto |1\rangle_b\langle 1| + O(|\xi|^2)\; .
\end{equation}
The fidelity of this process is high: $F=\langle 1|\rho|1\rangle\simeq 1$,
and this is therefore typically a very good single-photon state preparation 
process (although the situation changes drastically when multiple 
down-converters are considered \cite{braunstein98,kok00a}). Due to the large 
vacuum contribution, however, the probability of the detector giving a `click' 
will be small (of order $O(|\xi|^2)$). When the detector does not click, that 
particular trial is dismissed, hence the {\em conditional} character of the 
detection. 

In this example the outcome of the detection is used to either accept or 
reject a particular run of the state preparation device. However, in general 
the outcome of the detector can be used to determine a more complicated 
operation on the remainder of the state preparation process. This is detection
plus {\em feed-forward}, since the outcome is used further on in the 
process. An example of this is quantum teleportation, where the outcome of 
the Bell measurement determines the unitary transformation needed to retrieve 
the original input state.

When the measurements in the state preparation process are prone to errors,
the state we want to create may not be the state we actually create. This 
means that errors in the detection devices can lead to reduced fidelities. In
this paper we study the effect of detection errors on optical travelling-wave 
state preparation.

\section{Confidence}\label{confidence}

Consider a preparation device which prepares a state conditioned on a single 
measurement. For simplicity, we employ two subsystems. One subsystem will be 
measured, leaving a quantum state in the other. It is clear that prior to
the measurement the two systems have to be entangled. Otherwise conditioning
on the measurement does not have any effect on the state of the second system.

\begin{figure}[t]
  \begin{center}
  \begin{psfrags}
     \psfrag{a}{$a_k$}
     \psfrag{rho}{$\rho_{a_k}$}
     \psfrag{psi}{$|\psi\rangle$}
     \epsfxsize=8in
     \epsfbox[-100 20 800 120]{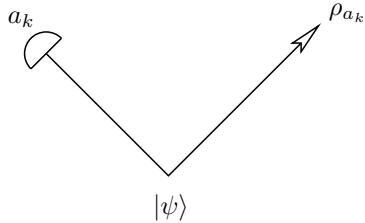}
  \end{psfrags}
  \end{center}
  \narrowtext
  \caption{A schematic representation of state preparation conditioned on a 
	measurement. One branch of the entanglement $|\psi\rangle$ is detected,
	yielding an eigenvalue $a_k$. The other branch is now in a state 
	$\rho_{a_k}$.}
  \label{fig:4.0}
\end{figure}

We can write the total state $|\psi\rangle_{12}$ prior to the measurement in
the Schmidt decomposition:
\begin{equation}\label{4schmidt}
 |\psi\rangle_{12} = \sum_k c_k |a_k\rangle_1 |b_k\rangle_2\; ,
\end{equation}
with $\{ |a_k\rangle\}_k$ and $\{ |b_k\rangle\}_k$ orthonormal sets of states 
for system 1 and 2 respectively. These states correspond to eigenstates of 
observables $A$ and $B$ with sets of eigenvalues $\{ a_k\}_k$ and $\{ b_k\}_k$ 
respectively. We now measure the observable $A$ in system 1, yielding an 
outcome $a_k$ (see Fig.\ \ref{fig:4.0}).

We can model this measurement using so-called projection operator valued
measures, or POVM's for short. For {\em ideal} measurements, we can describe 
the measurement of mode 1 as a projection $P_k = |a_k\rangle\langle a_k|$ 
operating on the state $|\psi\rangle_{12}$. When we trace out the first system 
the (normalised) state of the second system will be
\begin{equation}
 \rho_{a_k} = \frac{{\rm Tr}_1 [(P_k\otimes{\unity})|\psi\rangle_{12}
 \langle\psi|]}{{\rm Tr}_{12} [(P_k\otimes{\unity})|\psi\rangle_{12}
 \langle\psi|]} = |b_k\rangle\langle b_k|\; .
\end{equation}

For non-ideal measurements we do not use a projection operator, but rather 
a {\em projection operator valued measure}. In general, a POVM $E_{\nu}$
can be written as \cite{kraus83,helstrom76}
\begin{equation}\label{povm1}
 E_{\nu} = \sum_{\mu} d_{\mu\nu} {\mathcal{P}}_{\mu} \geq 0\; ,
\end{equation}
where the ${\mathcal{P}}_{\mu}$'s form a set (possibly over-complete, hence 
the difference in notation from $P_k$) of projection operators 
$\{ |\mu\rangle\langle\mu|\}_{\mu}$. We also require a completeness relation
\begin{equation}
 \sum_{\nu} E_{\nu} = \unity\; .
\end{equation}

As mentioned before, a measurement outcome $a_k$ in mode 1 gives rise to 
an outgoing state $\rho_{a_k}$ in mode 2. We cannot describe a non-ideal
measurement with the projection $P_k = |a_k\rangle\langle a_k|$. Instead, we 
have a POVM $E_k$ (corresponding to the outcome $a_k$), which reduces to $P_k$ 
in the case of an ideal measurement. Let $\rho_{12} = |\psi\rangle_{12}\langle
\psi|$, the entangled state prior to the measurement. The outgoing state in
mode $b$ will then be 
\begin{equation}\label{4rhoout}
 \rho_{a_k} = \frac{{\rm Tr}_1[(E_k\otimes{\unity})\rho_{12}]}{{\rm 
 Tr}[(E_k\otimes{\unity})\rho_{12}]}\; ,
\end{equation}
where the total trace over both systems in the denominator gives the proper
normalisation.

If we had an ideal detector (corresponding to $E_k = |a_k\rangle\langle a_k|$),
the outgoing state would be $\rho_{a_k} = |b_k\rangle\langle b_k|$. However, 
with the general POVM $E_k$, this will not be the case. The resulting state 
will be different. In order to quantify the reliability of a state preparation 
process we introduce the {\em confidence} of a process.
\begin{description}
 \item[Definition:] The {\em confidence} in the preparation of a particular
	state is given by the fidelity of the preparation process.
\end{description}
That means that using Eqs.\ (\ref{4schmidt}) and (\ref{4rhoout}) the 
confidence $C$ is given by
\begin{equation}\label{conf1}
 C = \frac{{\rm Tr}[(E_k\otimes|b_k\rangle\langle b_k|)\rho_{12}]}{{\rm 
 Tr}[(E_k\otimes{\unity})\rho_{12}]} = \frac{|c_k|^2 \langle a_k| E_k 
 |a_k\rangle}{\sum_l |c_l|^2 \langle a_l| E_k |a_l\rangle}\; ,
\end{equation}
where the $|c_l|^2$ are the diagonal elements of the density matrix. In the 
context of measurement and state identification, the fidelity is a widely used
and well-studied concept \cite{massar95,massar99}. Since the confidence is
defined as the fidelity of the preparation process, these results also apply 
here.

We prefer the term confidence in this context, because it is reminiscent of 
the confidence in statistics \cite{deutsch65}. Statistical confidence denotes 
the probability that the value of a quantity lies within a fixed interval 
around the observed mean value. In this paper, we extend this meaning to the 
quantum mechanical case. It is the probability that the {\em prepared} state 
passes a projective test for the {\em expected} state in a single-shot
experiment.

The confidence $C$ in Eq.~(\ref{conf1}) can be interpreted as the probability 
of obtaining outcome $a_k$ from the `branch' containing $|a_k\rangle$ in 
Eq.~(\ref{4schmidt}) divided by the unconditional probability of obtaining 
outcome $a_k$. We will also call this the `confidence of state preparation'. 
This interpretation suggests that there does not need to be a second system to 
give the idea of confidence meaning. Suppose, for instance, that we have an 
`electron factory' which produces electrons with random spin. A Stern-Gerlach 
apparatus in the path of such an electron will make a spin measurement along a 
certain direction {\bf r}. Suppose we find that the electron has spin `up' 
along {\bf r}. Before this measurement the electron was in a state of random 
spin ($\rho_{\rm in}=\frac{1}{2}{|\uparrow\rangle\langle\uparrow|} +
\frac{1}{2}{|\downarrow\rangle\langle\downarrow|}$), and after the measurement 
the electron is in the `spin up' state ($\rho_{\rm out} = |\uparrow\rangle
\langle\uparrow|$). The state of the electron has {\em collapsed} into the 
`spin up' state. We will now investigate how we can define the confidence of 
the detection of a single system.

Formally, we can model state collapse by means of the super-operator 
$\hat{\mathcal{F}}_{a_k}$, where $a_k$ is again the outcome of the measurement 
of observable $A$ (`spin up' in the above example). In general, a 
super-operator yields a (non-normalised) mapping $\rho\rightarrow
\hat{\mathcal{F}}_{\mu}(\rho)$ (see Fuchs and Peres \cite{fuchs96b,peres00} and
references therein). When the eigenstate corresponding to $a_k$ is given by 
$|a_k\rangle$, we can define the confidence of this measurement as 
\begin{equation}
 C_{\rm m} = \frac{\langle a_k|\hat{\mathcal{F}}_{a_k}(\rho)|a_k\rangle}{{\rm 
 Tr}[\hat{\mathcal{F}}_{a_k}(\rho)]} = \frac{{\rm Tr}[\hat{\mathcal{F}}_{a_k}
 (\rho)|a_k\rangle\langle a_k|]}{{\rm Tr}[\hat{\mathcal{F}}_{a_k}(\rho)]}\; ,
\end{equation}
with ${\rm Tr}[\hat{\mathcal{F}}_{a_k}(\rho)]$ the proper normalisation. 
However, this expression depends strongly on the details of the family of 
operators ${\mathcal{A}}_{\mu\nu}$. This is a more complicated generalisation 
than the POVM's $E_k$. The confidence of state preparation, on the other hand, 
is a function of the POVM $E_k$. Furthermore, $C_{\rm m}$ will in general {\em 
not} be equal to the confidence of state preparation derived in Eq.\ 
(\ref{conf1}). 

In conclusion, there are two distinct versions of the confidence: the 
confidence of {\em measurement} and the confidence of {\em state preparation}.
Later in this paper we will use the concept of the confidence to make a
quantitative comparison between different detection devices. This suggests 
that we need to calculate the confidence of measurement with all its 
difficulties. One way to circumvent this problem is to calculate the the 
confidence of state preparation using a fixed state. Instead of concentrating 
on the state preparation process we now choose a standard input state and 
calculate the confidence for different types of measurement devices. One such 
choice might be the maximally entangled state 
\begin{equation}
 |\Psi\rangle_{12} = \frac{1}{\sqrt{N}} \sum_{k=0}^{N-1} |a_k,a_k\rangle\; .
\end{equation}
When $N\rightarrow\infty$, this is perhaps not the ideal choice and another 
state may be preferred. For any choice, the confidence offers a quantitative 
measure of performance for different types of measurement devices. 

\section{Optical detection devices}\label{odd}

Having set the stage for state preparation conditioned on measurement outcomes,
we will now restrict the remainder of this paper to optical implementations.
Let's consider the measurement of optical Fock states using photo-detectors. 
In order to classify different types of detectors we use the following 
terminology: a detector is said to have a {\em single-photon sensitivity} when 
it is sensitive enough to detect a single-photon wave-packet. This is the case
with, for example, the avalanche photo-detector. When a detector can 
distinguish between $n$- and $(n+1)$-photon wave-packets, it is said to have 
a {\em single-photon resolution}. 

Real detectors have a variety of characteristics. Most common detectors do not 
have single-photon resolution, although they can distinguish between a few and 
many photons. When small photon numbers are detected, however, these are  
single-photon sensitivity detectors to a good approximation. There are also 
single-photon resolution detectors \cite{kim99,takeuchi99}. Currently, these
detectors require demanding operating conditions. At this point we note that 
here, we only consider the detection of single modes. In practice, however,
detectors are multi-mode detectors. Since we are dealing with direct detection,
these other modes only contribute to the background noise, with the quantum
efficiency to the single mode being the key parameter.

\begin{figure}[t]
  \begin{center}
  \begin{psfrags}
     \psfrag{2n}{$N$}
     \psfrag{n2}{$N$}
     \psfrag{p}{$N$-port}
     \psfrag{<}{$\left\{ \mbox{\rule{0mm}{8.5mm}} \right. $}	
     \psfrag{>}{$\left. \mbox{\rule{0mm}{8.5mm}} \right\} $}
     \epsfxsize=8in
     \epsfbox[-100 20 1000 100]{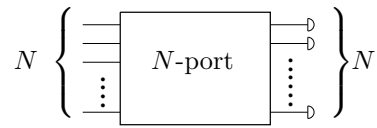}
  \end{psfrags}
  \end{center}
  \caption{An $N$-port with unit-efficiency, non-resolving detectors. The $N$ 
	incoming modes are unitarily transformed into $N$ output modes. The 
	$N$-ports considered here consist of mirrors and beam-splitters and do 
	not mix creation operators with annihilation operators.}
  \label{fig:4.1}
\end{figure}

When we need single-photon resolution but do not have the resources to employ
single-photon resolution detectors, we can use a so-called {\em detector 
cascade} \cite{song90}. In a detector cascade an incoming mode (populated by 
a number of photons) is split into $N$ output modes with equal amplitude which 
are all detected with single-photon sensitivity detectors. The idea is to 
choose the number of output modes large enough, so that the probability that 
two photons enter the same detector becomes small. In general, an optical 
setup which transforms $N$ incoming modes into $N$ outgoing modes is called an 
$N$-port (see Fig.\ \ref{fig:4.1}) \cite{reck94}. A detector cascade is a 
symmetric $N$-port with detectors at the outgoing modes and vacuum states in 
all input modes except the first mode. In the next section we will study the
statistics of symmetric $N$-ports, but first we need to elaborate on the types
of errors which occur in detectors.

There are two sources of errors for a detector: it might fail to detect a 
photon, or it might give a signal although there wasn't actually a photon 
present. The former may be characterised as a `detector loss' and the latter 
as a `dark count'. Here, the emphasis will be on detector losses. In some 
experiments (like the Innsbruck teleportation experiment \cite{bouwmeester97}) 
the detectors operate within short gated time intervals. This greatly reduces 
the effect of dark counts and we will not consider them here.

Detector losses are not so easily dismissed. Every photon entering a detector 
has a certain probability of triggering it. This probability is called the
{\em efficiency} of the detector. For the purposes of brevity, when a detector 
is perfectly efficient, we will call it a {\em unit-efficiency} detector. When 
it has some lower efficiency, we speak of a {\em finite-efficiency} detector.
Here, we study detector cascading with unit-efficiency detectors, as well
as cascading with finite-efficiency detectors \cite{yuen80}. we are interested 
in the case where cascading distinguishes between photon-number states 
$|k\rangle$ and $|k'\rangle$ with $k\simeq k'$. 

\section{{\it N}-ports}\label{sec:nports}

In this section we treat the properties of detector cascades, or symmetric 
$N$-ports with single-photon sensitivity detectors in the outgoing modes. 
Symmetric $N$-ports yield a (unitary) transformation $U$ of the spatial field
modes $a_k$, with $j,k=1,\ldots,N$:
\begin{equation}\label{trans1}
 \hat{b}_k \rightarrow \sum_{j=1}^N U_{jk} \hat{a}_j
 \qquad\mbox{and}\qquad
 \hat{b}^{\dagger}_k \rightarrow \sum_{j=1}^N U_{jk}^*\hat{a}^{\dagger}_j\; ,
\end{equation} 
where the incoming modes of the $N$-port are denoted by $a_j$ and the outgoing 
modes by $b_j$. Here, $\hat{a}^{\dagger}_j$ and $\hat{a}_j$ are the respective 
creation and annihilation operators of mode $a_j$. Similarly for mode $b_k$.
The unitary matrix $U$ can be chosen to be
\begin{equation}\label{trans2}
  U_{jk} = \frac{1}{\sqrt{N}}\exp[2\pi i (j-1)(k-1)/N]
\end{equation} 
without loss of generality up to an overall phase-factor. Paul {\em et al}.\ 
have studied such devices in the context of tomography and homodyne detection 
\cite{paul96}. 

Here, we study $N$-ports in the context of optical state preparation, where 
only one copy of a state is given, instead of an ensemble. we will use
the concept of the {\em confidence}, introduced in section {\ref{confidence}}.

\subsection{Statistics of {\it N}-ports}

Suppose we have a detector cascade, consisting of a symmetric $N$-port with 
single-photon sensitivity detectors in the outgoing modes. According to 
Eqs.\ (\ref{trans1}) and (\ref{trans2}) incoming photons will be redistributed 
over the outgoing modes. In this section we study the photon statistics of this 
device. In particular, we study the case where $k$ photons enter a single input 
mode of the $N$-port, with vacuum in all other input modes. This device (i.e., 
the detector cascade) will act as a sub-ideal single-photon resolution 
detector since there is a probability that some of the photons end up in the 
same outgoing mode, thus triggering the same detector.

To quantify the single-photon resolution of the cascade we use the confidence
given by Eq.\ (\ref{conf1}). Suppose we have two spatially separated entangled 
modes of the electro-magnetic field $a$ and $b$ with number states $|m\rangle$
in $a$ and some other orthogonal states $|\phi_m\rangle$ in $b$:
\begin{equation}
 |\Psi\rangle = \sum_m \gamma_m |m\rangle_a |\phi_m\rangle_b\; ,
\end{equation}
where the second mode is used only to give the confidence an operational 
meaning. The POVM governing the detection can be written as $E_k = \sum_m 
p_N (k|m) |m\rangle\langle m|$, since we assume that the photons are not 
lost in the $N$-port. In this expression $p_N(k|m)$ is the probability that 
$m$ incoming photons cause a $k$-fold detector coincidence in the $N$-port 
cascade. The confidence can then be written as
\begin{equation}\label{conf2}
 C = \frac{|\gamma_k|^2 \langle k|E_k|k\rangle}{\sum_l|\gamma_m|^2 \langle m|
 E_k|m\rangle} =\frac{|\gamma_k|^2 p_N (k|k)}{\sum_m|\gamma_m|^2 p_N (k|m)}\; .
\end{equation}
In order to find the confidence, we therefore first have to calculate the 
probability distribution $p_N$. This will allow us to compare single-photon 
resolution detectors with various arrangements ($N$-ports) of single-photon 
sensitivity detectors.

Suppose $k$ photons enter the first input mode and all other input modes are
in the vacuum state. The density matrix of the pure input state $\rho_0 = 
|k\rangle\langle k|$ will be transformed according to $\rho = U_N \rho_0 
U^{\dagger}_N$ with $U_N$ the unitary transformation associated with the 
symmetric $N$-port. Let $\vec{n}$ be the $N$-tuple of the photon number in 
every outgoing mode: $\vec{n}= (n_1,n_2,\ldots, n_N)$. The probability of 
finding $n_1$ photons in mode 1 and $n_2$ photons in mode 2, et cetera, is 
given by $p_{\vec{n}} = \langle\vec{n}|\rho|\vec{n}\rangle$. Using the 
$N$-port transformation this probability yields 
\begin{equation}
 p_{\vec{n}} = \langle\vec{n}|U_N\rho_0 U^{\dagger}_N|\vec{n}\rangle = 
 |\langle\vec{n}|U_N|\vec{k}\rangle|^2\; ,
\end{equation}
where $\vec{k}=(k,0,\ldots, 0)$, since only the first input mode inhabits
photons and the rest are vacuum. From Refs.\ \cite{dodonov94} and 
\cite{dodonov96} we find that this can be rewritten as 
\begin{equation}\label{prob}
 p_{\vec{n}} = \frac{\left[ H^R_{\vec{k}\vec{n}}(0)\right]^2}{n_1!
  \cdots n_N! k!}\; .
\end{equation}
Here, $H^R_{\vec{k}\vec{n}}(\vec{x})$ is a so-called multi-dimensional 
Hermite polynomial (MDHP) \cite{dodonov84} and the matrix $R$ is defined as
\begin{equation}\label{matrix}
 R \equiv
 \begin{pmatrix}
  0 & -U^{\dagger} \cr -U^{\dagger} & 0
 \end{pmatrix}\; .
\end{equation}
For our present purposes it is convenient to characterise the $N$-port by 
its trans\-formation of the field modes given by Eqs.\ (\ref{trans1}) and 
(\ref{trans2}). we therefore concentrate on $U$ rather than $U_N$.

Since there is a one-to-one correspondence between the $N$-port ($U$) and the 
matrix $R$, knowledge of $U$ is sufficient to calculate the confidence 
of a given event using the $N$-port. The MDHP for $N$ input modes with $k$ 
photons in the first mode and zero in the others (giving an $N$-tuple 
$\vec{k}$) and $N$ output modes $\vec{n}$ is given by
\begin{equation}
 H_{\vec{k}\vec{n}}^R (\vec{x}) = (-1)^{2k}\; e^{\frac{1}{2}
 \vec{x}\,R\,\vec{x}^T}\; \nabla_{\vec{k}\vec{n}}^{2k}\; e^{-\frac{1}{2}\vec{x}
 \,R\,\vec{x}^T} \; ,
\end{equation}
where $\vec{x}\,R\,\vec{x}^T = \sum_{ij} x_i R_{ij} x_j$, $\vec{x} = (x_1,
\ldots,x_{2N})$ and
\begin{equation}\nonumber
 \nabla_{\vec{k}\vec{n}}^{2k} \equiv \frac{\partial^{2k}}{\partial x^k_1 
  \partial x^{n_1}_{N+1} \cdots \partial x^{n_N}_{2N}}\; .
\end{equation}
The number of photons in the input mode is equal to the total number of 
photons in the output modes. The dimension of $\vec{x}$ obeys $\dim\vec{x} = 
\dim \vec{k} + \dim\vec{n} = 2N$. For example, for a two-photon input state we 
have
\begin{equation}\label{2photon}
 e^{\frac{1}{2}\vec{x}\,R\,\vec{x}^T} \frac{\partial^4}{\partial x_1^2\partial 
 x_l \partial x_k} \left. 
 e^{-\frac{1}{2}\vec{x}\,R\,\vec{x}^T} \right|_{\vec{x}=0} = 2R_{1l}R_{1k}\; .
\end{equation}

There are many different ways in which $k$ incoming photons can trigger a 
$k$-fold detector coincidence. These different ways correspond to different 
photon distributions in the outgoing (detected) modes, and are labelled by 
$\vec{n}_r$. The probability that all $k$ photons enter a different 
detector is found by determining the $p_{\vec{n}_r}$s where every $n_i$ in 
$\vec{n}_r$ is at most one. The sum over all these $p_{\vec{n}_r}$'s is equal 
to the probability $p_N (k|k)$ of a $k$-fold coincidence in an $N$-port 
conditioned on $k$ incoming photons:
\begin{equation}\label{pkn}
 p_N(k|k) = \sum_{\vec{n}_r} p_{\vec{n}_r} = \frac{k!}{N^k} \binom{N}{k}\; .
\end{equation}

Finally, in order to find the probability of a $k$-fold detector coincidence 
conditioned on $m$ photons in the input state (with $m\geq k$) we need to sum 
all probabilities in Eq.\ (\ref{prob}) with $k$ non-zero entries in the 
$N$-tuple $\vec{n}$:
\begin{equation}\label{pkm1}
 p_N (k|m) = \sum_{\vec{n}\in{\mathcal{S}}_k }\frac{\left[ H^R_{\vec{m}
 \vec{n}}(0)\right]^2}{n_1!\cdots n_N! m!}\; ,
\end{equation}
where ${\mathcal{S}}_k$ is the set of all $\vec{n}$ with exactly $k$ non-zero
entries. 

\subsection{Realistic {\it N}-ports}{\label{realisticnports}}

\begin{figure}[t]
  \begin{center}
  \begin{psfrags}
     \psfrag{2n}{$2N$}	
     \psfrag{n1}{$N$}	
     \psfrag{n2}{$N$}	
     \psfrag{p}{$2N$-port}	
     \psfrag{<}{$\left\{ \mbox{\rule{0mm}{9mm}} \right. $}	
     \psfrag{>}{$\left. \mbox{\rule{0mm}{9mm}} \right\} $}
     \psfrag{2}{$\overbrace{\phantom{xxxxxx}}$}	
     \epsfxsize=8in
     \epsfbox[-80 20 940 140]{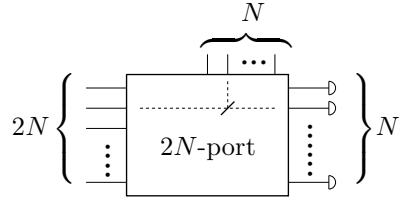}
  \end{psfrags}
  \end{center}
  \caption{A $2N$-port with $N$ modes which are detected with ideal
	detectors and $N$ undetected modes. These modes are associated with 
	the detector losses.}
  \label{fig:4.2}
\end{figure}

We now consider a symmetric $N$-port cascade with finite-efficiency 
single-photon sensitivity
detectors. Every one of the $N$ detectors has a certain loss, which means that
some photons do not trigger the detector they enter. We can model this 
situation by putting a beam-splitter with intensity transmission coefficient 
$\eta^2$ in front of the ideal detectors \cite{yuen80}. The reflected photons 
are sent into the environment and can be associated with the loss. The 
transmitted photons are detected (see Fig.\ \ref{fig:4.2}). 

The implementation of the beam-splitters responsible for the detector losses 
transform our $N$-port into a $2N$-port and the unitary transformation $U$ 
of the field modes in this $N$-port now becomes a $2N\times 2N$ unitary matrix 
$U~\rightarrow~ U\otimes {\unity}_2$ (where ${\unity}_2$ is the two dimensional
unit matrix). Applying a transformation $V_{\eta}$ to implement the 
beam-splitters with transmission coefficient $\eta^2$ will give a new unitary 
transformation governing the behaviour of the $2N$-port. Although nothing 
holds us from considering detectors with different efficiencies, for 
simplicity we will assume that all detectors have the same efficiency $\eta^2$.
In terms of the original unitary matrix $U$ from Eq.\ (\ref{trans2}) the new 
unitary matrix $\widetilde{U}$ becomes 
\begin{equation}\nonumber
 U ~\rightarrow~ \widetilde{U} =
 \begin{pmatrix}
  \eta\; U & \sqrt{1-\eta^2}\; U \cr -\sqrt{1-\eta^2}\; U & \eta\; U
 \end{pmatrix}\; .
\end{equation} 
This changes the matrix $R$ of the MDHP accordingly:
\begin{equation}
 R ~\rightarrow~ \widetilde{R} =
 \begin{pmatrix}
  0 & -\widetilde{U}^{\dagger} \cr -\widetilde{U}^{\dagger} & 0
 \end{pmatrix}
\end{equation}
and $\widetilde{R}$ is now a $4N \times 4N$ matrix dependent on $\eta$. The 
probability of finding a $k$-fold detector coincidence in an $N$-port cascade
with finite-efficient detectors then becomes
\begin{equation}\label{pkm}
 p_{N}(k|m) = \sum_{\vec{n}\in{\mathcal{S}}_k}\frac{\left[ H_{\vec{m}\vec{n}}^{
 \widetilde{R}}(0)\right]^2 }{n_1! \cdots n_{2N}!m!} \; ,
\end{equation}
where ${\mathcal{S}}_k$ is the set of all $\vec{n}$ with exactly $k$ non-zero
entries in the detected modes (note that we still call it an $N$-port although 
technically it is a $2N$-port). The confidence of having a total of $k$ photons
in a $k$-fold detector coincidence is again given by Eq.\ (\ref{conf2}).
The variables of the MDHP will be a $2N$-tuple $\vec{k} = (k, 0, \ldots 
0)$. The output photon number $2N$-tuple can now be written as $\vec{n} = 
(n^d_1, n^d_2,\ldots n^d_N, n_1^u, \ldots n^u_N)$, where the superscripts $d$ 
and $u$ again denote the detected and undetected modes respectively. 
Furthermore we have $\sum_{i=1}^N n^d_i \equiv N_d$ and $\sum_{i=1}^N n^u_i 
\equiv N_u$.

Using Eq.\ (\ref{pkn}) and observing that every detected photon carries a 
factor $\eta^2$ it is quite straightforward to obtain the probability that $k$ 
photons give a $k$-fold coincidence in an efficient $N$-port cascade:
\begin{equation}
 p_{N}(k|k) = \frac{\eta^{2k} N!}{N^k (N-k)!}\; .
\end{equation}

\subsection{The single-photon resolution of {\it N}-ports}

Having determined the probability distribution $p_N$, we can now calculate 
the confidence of detector cascading. First of all, in order to obtain a high 
confidence in the outcome of a detector cascade, the possible number of photons
should be much smaller than the number of modes in the cascade: $N\gg k$. In 
practice there is a limit to the number of detectors we can build a cascade 
with, so we only look at the lowest order: distinguishing between one and two 
photons.

we will calculate the confidence of having outgoing state $|\phi_1\rangle$ 
conditioned a single detector giving a `click' in the detector cascade when 
the input state is given by
\begin{equation}\label{confinput}
 |\Psi\rangle_{12} = \alpha |0\rangle_1 |\phi_0\rangle_2 + 
 \beta|1\rangle_1|\phi_1\rangle_2 + \gamma|2\rangle_1|\phi_2\rangle_2\; .
\end{equation}
This state corresponds, for example, to the output of a down-converter when we
ignore higher-order terms. The confidence is then
\begin{equation}
 C (1,|\Psi\rangle_{12}) = \frac{|\beta|^2 p_{N}(1|1)}{|\alpha|^2 p_N(1|0) + 
 |\beta|^2p_{N}(1|1) + |\gamma|^2 p_{N}(1|2)}\; .
\end{equation}
Eqs.\ (\ref{pkm}) and (\ref{2photon}) allow us to calculate the probabilities
of a zero-, one- and two-fold detector coincidence conditioned on one or two
incoming photons:
\begin{mathletters}
\begin{eqnarray}
 p_N (0|0) &=& 1 \\
 p_N (1|0) &=& 0 \\ && \cr
 p_N (0|1) &=& 1-\eta^2 \\ 
 p_N (1|1) &=& \eta^2 \\ && \cr
 p_N (0|2) &=& (1-\eta^2)^2 \\
 p_N (1|2) &=& \frac{\eta^4}{N} + 2\eta^2 (1-\eta^2) \label{fout} \\ 
 p_N (2|2) &=& \frac{N-1}{N} \eta^4\; ,
\end{eqnarray}
\end{mathletters}
For example, using these probabilities, together with Eq.\ (\ref{confinput}), 
gives us an expression for the confidence that a single detector hit was 
triggered by one photon ($\delta = |\gamma|^2 / |\beta|^2$):
\begin{equation}\label{2con2}
 C = \frac{N}{N + \delta[\eta^2 + 2N(1-\eta^2)]}\; ,
\end{equation}
where, for simplicity, we omitted the functional dependence of $C$ on the 
incoming state, the size of the cascade and the order of the detector 
coincidence. This gives a general measure of performance of an cascade of {\em 
arbitrary size} $N$ for the detection of up to two photons. Since the size
of the cascade needs to be comfortably larger than the number of detected 
photons, Eq.\ (\ref{2con2}) will be sufficient for most practical purposes.

A close look at Eq. (\ref{fout}) shows us that $p_N (1|2)$ includes a term
which is independent of the number of modes in the $N$-port cascade. This term
takes on a maximum value of $1/2$ for $\eta^2=\frac{1}{2}$. However, the 
confidence is a monotonously increasing function of $\eta^2$. As expected, 
for small $\delta$'s the confidence $C_N (1,|\Psi\rangle)$ approaches 1.
Detector cascading thus turns a collection of single-photon sensitivity
detectors into a device with {\em some} single-photon resolution. In the next
section we will give a quantitative estimation of this resolution.

\section{Comparing detection devices}\label{compare}

Let's return again to the schematic state preparation process depicted in 
figure \ref{fig:4.0}. There we had two modes, one of which was detected, giving
the prepared outgoing state in the other. we argued that different detection 
devices yield different output states, and the comparison of these states with 
the ideal case (where we used an ideal detector) led to the introduction of the
confidence of a state preparation process. Here, we will use the confidence
to make a comparison of different {\em detection devices}, rather than output
states. This can be done by choosing a fixed entangled input state. The 
confidence then quantifies the performance of these detection devices.

Consider the state preparation process in the setting of quantum optics. We
have two spatial modes of the electro-magnetic field, one of which is detected.
In this paper we are mostly interested in states containing a few photons, and
the detection devices we consider therefore include single-photon sensitivity
detectors, single-photon resolution detectors and detector cascades. As an 
example, we set the task of distinguishing between one and two photons. Since
single-photon sensitivity detectors are not capable of doing this, we will
compare the performance of detector cascading with that of a single-photon
resolution detector. Let the state prior to the detection be given by 
\begin{equation}
 |\Psi\rangle = \frac{1}{\sqrt{3}} \left( |0\rangle |\phi_0\rangle + 
 |1\rangle |\phi_1\rangle + |2\rangle |\phi_2\rangle \right)\; .
\end{equation}
This state is maximally entangled and will serve as our `benchmark' state. It
leads to the choice $\delta=1$ in Eq.\ (\ref{2con2}) in the previous section. 
Suppose the outgoing state conditioned on a `one-photon' indication in the 
detection device is $\rho$. The confidence is then again given by $C = 
\langle\phi_1|\rho|\phi_1\rangle$.

First, consider the single-photon resolution detector described in Refs.\ 
\cite{kim99,takeuchi99}. This detector can distinguish between one and two
photons very well, but it does suffer from detector losses (the efficiency was
determined at 88\%). That means that a two-photon state can be identified as 
a single-photon state when one photon is lost. The confidence of this detector 
is therefore not perfect.

In order to model the finite efficiency of the single-photon resolution 
detector we employ the beam-splitter model from section 
{\ref{realisticnports}}. We write the input state as
\begin{equation}
 |\Psi\rangle = \frac{1}{\sqrt{3}} \left( |0\rangle|\phi_0\rangle + 
 \hat{a}^{\dagger} |0\rangle |\phi_1\rangle + \frac{(\hat{a}^{\dagger})^2}
 {\sqrt{2}} |0\rangle |\phi_2\rangle \right)\; .
\end{equation}
When we make the substitution $\hat{a}^{\dagger}\rightarrow \eta 
\hat{b}^{\dagger} + \sqrt{1-\eta^2}\hat{c}^{\dagger}$ we obtain a state $\rho$.
The outgoing density matrix conditioned on a single photon in mode $b$ is then
\begin{eqnarray}
 \rho_{\rm out} &=& \frac{{\rm Tr}_{bc}[(|1\rangle_b\langle 1|\otimes
 {\unity}_c)\rho]}{{\rm Tr}[(|1\rangle_b\langle 1|\otimes{\unity}_c)\rho]}\cr 
 &=& \frac{\eta^2}{4-3\eta^2} |\phi_1\rangle\langle\phi_1| + \frac{4(1-\eta^2)}
 {4-3\eta^2} |\phi_2\rangle\langle\phi_2|\; .
\end{eqnarray}
With $\eta^2=0.88$ the confidence of the single-photon resolution detector is 
easily calculated to be $C=0.65$.

\begin{figure}[t]
  \begin{center}
  \begin{psfrags}
     \psfrag{0.2}{\footnotesize 0.2}	
     \psfrag{0.4}{\footnotesize 0.4}	
     \psfrag{0.6}{\footnotesize 0.6}	
     \psfrag{0.8}{\footnotesize 0.8}	
     \psfrag{1}{\footnotesize 1}	
     \psfrag{0}{\footnotesize 0}	
     \psfrag{e}{$\!\!\eta^2$}
     \psfrag{c}{$\!\!\!\!C$}
     \psfrag{a}{\small $N=1$}
     \psfrag{b}{\small $N=4$}
     \psfrag{f}{\small $N=16$}
     \psfrag{g}{\small $N=\infty$}
     \psfrag{h}{$\!\!\!\!\!\!\!\!\!\!\!\!\! |\Psi\rangle = 
	(|0\rangle|\phi_0\rangle + |1\rangle|\phi_1\rangle + 
	|2\rangle|\phi_2\rangle)/\sqrt{3}$}
     \epsfxsize=8in
     \epsfbox[0 70 700 230]{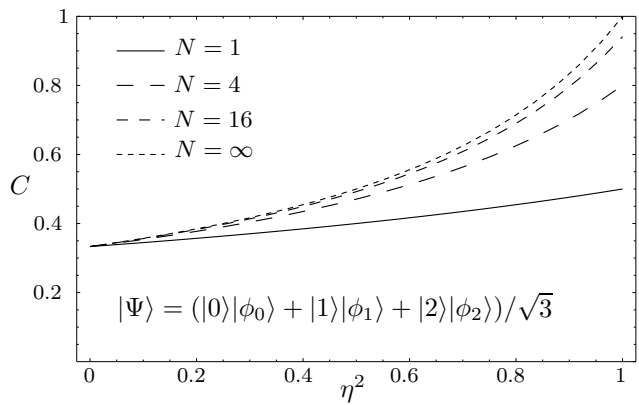}
  \end{psfrags}
  \end{center}
  \caption{The single-photon confidence $C$ [Eq.\ (\ref{2con2})] as 
	a function of the detector efficiency $\eta^2$. The solid line 
	corresponds to a single-detector cascade (no cascading: $N=1$), the 
	dashed lines correspond to $N=4$, $N=16$ and $N=\infty$ in ascending 
	order. We consider a maximally entangled input state $|\Psi\rangle = 
	(|0\rangle|\phi_0\rangle + |1\rangle|\phi_1\rangle + 
	|2\rangle|\phi_2\rangle)/\sqrt{3}$ to serve as a benchmark.}
  \label{fig:4.3}
\end{figure}

Now we consider a detector cascade with single-photon sensitivity detectors. 
In Fig.\ \ref{fig:4.3} the confidence of a single-photon detection with 
$N$-port cascades is depicted. When the cascade consists of four detectors 
($N=4$) it can be easily calculated from Eq.\ (\ref{2con2}) that the detectors 
need an efficiency of 0.84 to achieve a 0.65 confidence. In the case of 
infinite cascading ($N=\infty$) the single-photon confidence of 0.65 is met 
only if the efficiency is roughly 0.73. This puts a severe practical limit on 
the efficiency of the single-photon sensitivity detectors in the cascade. 

Detector cascading would be practically useful if a reasonably small 
number of finite-efficiency detectors yields a high confidence. In particular 
when cascading is viewed as an economical alternative to a detector with 
single-photon resolution the number of detectors in the cascade should be 
small. Additionally, cascading should yield a confidence similar to 
single-photon resolution detectors. Unfortunately, as a practical application, 
detector cascading only appears to yield a modest boost in resolution, unless 
the detectors with single-photon sensitivity have a very high efficiency. In
the context of entanglement-based state preparation, real single-photon 
resolution detectors are therefore superior to detector cascading with 
currently available detectors, notwithstanding the demanding operating 
conditions. 

\section{Conclusions}

In this paper we have studied the use of detection devices in 
entanglement-based travelling-wave state preparation. In particular we 
considered optical devices such as single-photon sensitivity detectors, 
single-photon resolution detectors and detector cascades. 

Detector cascading has generally been regarded as a good way to enhance 
single-photon resolution and consequently the fidelity of a state preparation
process \cite{song90}. However, an extensive theory for the use of these
detection devices has not been available so far. The statistics of $N$-ports
have been considered in the context of tomography \cite{paul96}, which relies 
on the availability of a large number of copies of a quantum state. In
state preparation, however, we perform measurements on single systems, and we 
therefore need precise bounds on the distinguishability of these measurements.

To this end, we introduced the confidence of preparation, which can also be 
used to quantify the (preparation) performance of a (realistic) detection 
device. We gave an expression for the confidence of a cascade of arbitrary 
size $N$, conditioned on an input state of up to two photons. We believe that 
this will be sufficient for most practical purposes. Thus, we compared a 
single-photon resolution detector with a cascade of single-photon sensitivity 
detectors and found that cascading {\em does not give a practical advantage} 
over detectors with single-photon resolution for entanglement-based state 
preparation.

This research was funded in part by EPSRC grant GR/L91344.

\end{multicols}

\end{document}